\documentclass[10pt,twocolumn]{IEEEtran}

\makeatletter
\def\ifundefined{\@ifundefined}
\makeatother

\begin{document}

\title{Improved Upper Bound for the Redundancy of Fix-Free Codes}

\author{Sergey Yekhanin\thanks{The author is a Ph.D. student at the
Department of Electrical Engineering and Computer Science,
Massachusetts Institute of Technology, 
Cambridge, MA 02139, USA ~(email: yekhanin@mit.edu)}}

\ifundefined{IEEEtransversionmajor}{%
   \newlength{\IEEEilabelindent}
   \newlength{\IEEEilabelindentA}
   \newlength{\IEEEilabelindentB}
   \newlength{\IEEEelabelindent}
   \newlength{\IEEEdlabelindent}
   \newlength{\labelindent}
   \newlength{\IEEEiednormlabelsep}
   \newlength{\IEEEiedmathlabelsep}
   \newlength{\IEEEiedtopsep}

   \providecommand{\IEEElabelindentfactori}{1.0}
   \providecommand{\IEEElabelindentfactorii}{0.75}
   \providecommand{\IEEElabelindentfactoriii}{0.0}
   \providecommand{\IEEElabelindentfactoriv}{0.0}
   \providecommand{\IEEElabelindentfactorv}{0.0}
   \providecommand{\IEEElabelindentfactorvi}{0.0}
   \providecommand{\labelindentfactor}{1.0}
   
   \providecommand{\iedlistdecl}{\relax}
   \providecommand{\calcleftmargin}[1]{
                   \setlength{\leftmargin}{#1}
                   \addtolength{\leftmargin}{\labelwidth}
                   \addtolength{\leftmargin}{\labelsep}}
   \providecommand{\setlabelwidth}[1]{
                   \settowidth{\labelwidth}{#1}} 
   \providecommand{\usemathlabelsep}{\relax}
   \providecommand{\iedlabeljustifyl}{\relax}
   \providecommand{\iedlabeljustifyc}{\relax}
   \providecommand{\iedlabeljustifyr}{\relax}
 
   \newif\ifnocalcleftmargin
   \nocalcleftmarginfalse

   \newif\ifnolabelindentfactor
   \nolabelindentfactorfalse
   
   \newif\ifcenterfigcaptions
   \centerfigcaptionsfalse
   
   \let\OLDitemize\itemize
   \let\OLDenumerate\enumerate
   \let\OLDdescription\description
   
   \renewcommand{\itemize}[1][\relax]{\OLDitemize}
   \renewcommand{\enumerate}[1][\relax]{\OLDenumerate}
   \renewcommand{\description}[1][\relax]{\OLDdescription}

   \providecommand{\pubid}[1]{\relax}
   \providecommand{\pubidadjcol}{\relax}
   \providecommand{\specialpapernotice}[1]{\relax}
   \providecommand{\overrideIEEEmargins}{\relax}
   
   \let\CMPARstart\PARstart 
   
   \let\OLDappendix\appendix
   \renewcommand{\appendix}[1][\relax]{\OLDappendix}
   
   \newif\ifuseRomanappendices
   \useRomanappendicestrue
   
   \let\OLDbiography\biography
   \let\OLDendbiography\endbiography
   \renewcommand{\biography}[2][\relax]{\OLDbiography{#2}}
   \renewcommand{\endbiography}{\OLDendbiography}
   
   \markboth{ {}}{}}{
  
   \markboth{ }}%
   {Shell: A Test for IEEEtran.cls}

\renewcommand{\PARstart}[2]{\CMPARstart{#1}{#2}}

\maketitle

\begin{abstract}
A variable-length code is a fix-free code if no codeword is a 
prefix or a suffix of any other codeword. In a fix-free code any
finite sequence of codewords can be decoded in both directions,
which can improve the robustness to channel noise and speed up the 
decoding process. In this paper we prove a new sufficient 
condition of the existence of fix-free codes and improve the upper
bound on the redundancy of optimal fix-free codes.
\end{abstract}



\section{Introduction}

\PARstart{L}{et} $p=\{p_1,\ldots,p_m\}$ be the probability distribution
of a source, and let $C$ be a code for the source. The redundancy 
$R$ of a code $C$ is defined as the difference between the average
codeword length $L(C)$ of this code and the entropy $H(p)$ of the 
source. We denote the redundancy of an optimal fix-free code by $R_f$.

Ahlswede {\it et al.} \cite{Ahlswede} have proved that $0\leq R_f <2$.
They have also shown that the lower bound $0$ on $R_f$ cannot be
improved. Later Ye and Yeung \cite{Ye_Yeung_conf,Ye_Yeung_paper} 
derived several upper bounds on $R_f$ in terms of partial information
about the source distribution. The goal of this paper is to improve the
upper bound on $R_f$ from $2$ to $4-\log_{2}5$, which is approximately
$1.678$.

Let $\mathbf{v}_n=(k_1,\ldots,k_n)$ be a vector, where $k_i$ are 
nonnegative integers. By $C(\mathbf{v}_n)$ denote a binary 
variable-length code containing $k_i$ codewords of length
$i$, for each $i=\overline{1,n}$. The Kraft sum of the vector 
$\mathbf{v}_n$
is the quantity
\begin{equation}\label{s_def}
S(\mathbf{v}_n)=\sum\limits_{i=1}^{n}\frac{k_i}{2^i}.
\end{equation}

Ahlswede {\it et al.} \cite{Ahlswede} conjectured that 
$S(\mathbf{v}_n)\leq\frac{3}{4}$ is a sufficient condition for the
existence
of a binary fix-free code $C(\mathbf{v}_n)$. They proved 
that the conjecture
is true in the weaker case when the Kraft sum is
at most $\frac{1}{2}$.
If the conjecture is true the upper bound on 
$R_f$ can be improved
to $3-\log_{2}3$, which is approximately 
$1.415$~\cite{Ye_Yeung_paper}.
Since the conjecture was made, 
several special cases of it were proven 
\cite{Harada_Kobayashi,Kukorelly_Zeger,Ye_Yeung_paper,Yekhanin},
although the general conjecture still remains an open problem.

In this paper we prove a new special case of the conjecture.
We show that $S(\mathbf{v}_n)\leq\frac{5}{8}$ implies the existence 
of a fix-free code $C(\mathbf{v}_n)$~(Theorem 1). This result yields
an improved upper bound for the redundancy of optimal fix-free
codes (Theorem 2).

\section{New Sufficient Condition of the Existence \\ of Fix-Free Codes}

Let $\mathbf{w}$ be an arbitrary binary vector of length $n$.
The binary vector composed of the first \{last\} $p$ symbols of $\mathbf{w}$
is called $p$-{\it prefix} \{$p$-{\it suffix}\} of $\mathbf{w}$ and denoted by
$^p\mathbf{w}$ \{$\mathbf{w}^p$\}. 
We say that vector $\mathbf{w}$ has the form $\alpha\star\beta$, where 
$\alpha, \beta \in \{0,1\}$ if $^1\mathbf{w}=\alpha$ and $\mathbf{w}^1=\beta$.

Consider a binary variable-length fix-free code $C(\mathbf{v}_n)$, where 
$\mathbf{v}_n=(k_1,\ldots,k_n)$.

A vector $\mathbf{w} \in \{0,1\}^n$ is called 
{\it prefix free} \{{\it suffix free}\} over code $C(\mathbf{v}_n)$ if
$C(\mathbf{v}_n)$ does not contain any prefix \{suffix\} of $\mathbf{w}$.

By definition, put
$$
\begin{array}{c}
^0\overrightarrow {F}(C)=\{\mathbf{w} | \mathbf{w} 
    \ \mbox{is prefix-free over} \ C \ \mbox{and}\  ^1\mathbf{w}=0\}, \\

^1\overrightarrow {F}(C)=\{\mathbf{w} | \mathbf{w} 
    \ \mbox{is prefix-free over} \ C \ \mbox{and}\  ^1\mathbf{w}=1\}, \\

\overrightarrow {F}(C)=^0\overrightarrow {F}(C) \cup ^1\overrightarrow {F}(C), \\

 \overleftarrow {F}^0(C)=\{\mathbf{w} | \mathbf{w} 
    \ \mbox{is suffix-free over} \ C \ \mbox{and}\  \mathbf{w}^1=0\}, \\

 \overleftarrow {F}^1(C)=\{\mathbf{w} | \mathbf{w} 
    \ \mbox{is suffix-free over} \ C \ \mbox{and}\  \mathbf{w}^1=1\}, \\

\overleftarrow {F}(C)= \overleftarrow {F}^0(C) \cup  \overleftarrow {F}^1(C).
 
\end{array}
$$

Let $M$ be an arbitrary subset of $\{0,1\}^n$. 

The set $M$ 
is called {\it right regular} if all $(n-1)$-suffixes of words from $M$
are pairwise distinct, i.e., $\forall c_1,c_2 \in M,\  c_1\neq c_2$
implies $c_1^{n-1}\neq c_2^{n-1}$. 

Similarly, the set $M$ is called {\it left regular}
if all $(n-1)$-prefixes of words from $M$ are pairwise distinct,
i.e., $\forall c_1,c_2 \in M,\  c_1\neq c_2$ implies 
$^{n-1}c_1\neq ^{n-1}c_2$.

Clearly, $^0\overrightarrow {F}(C)$ and $^1\overrightarrow {F}(C)$ are right
regular sets. Likewise, $\overleftarrow {F}^0(C)$ and $\overleftarrow {F}^1(C)$
are left regular sets.

Let $M_1$ and $M_2$ be arbitrary subsets of $\{0,1\}^n$.
By definition, put
$$
M_1\otimes M_2=\{\mathbf{w}\in \{0,1\}^{n+1}\ | 
   ^{n}\mathbf{w}\in M_1 \ \mbox{and} \ \mathbf{w}^n \in M_2\}
$$

The following lemma is obvious.

{\bf Lemma 1:}
Suppose $C(\mathbf{v}_n)$ is an arbitrary fix-free code;
then $\overrightarrow{F}(C)\otimes \overleftarrow{F}(C)$ 
is the set of all words of length $n+1$ that can be added to $C(\mathbf{v}_n)$
without violation of the fix-free property of the code. Moreover,
$^\alpha\overrightarrow{F}(C)\otimes \overleftarrow{F}^\beta (C)$ 
is the set of all words of the form $\alpha\star\beta$ and length $n+1$ 
that can be added to $C(\mathbf{v}_n)$ without violation of the fix-free
property of the code.

{\bf Lemma 2:} Suppose $M_1$ is a right regular subset of $\{0,1\}^n$ 
and $M_2$ is a left regular subset of $\{0,1\}^n$; then
\begin{equation}\label{cross_volume}
|M_1\otimes M_2|\geq |M_1|+|M_2|-2^{n-1}. 
\end{equation}

{\bf Proof:} By $M_1^{(n-1)}$ denote the set of $(n-1)$-suffixes of 
words from $M_1$. In the same way, by $^{(n-1)}M_2$ denote the set of
$(n-1)$-prefixes of words from $M_2$. Since $M_1$ is right regular, 
it follows that $|M_1^{(n-1)}|=|M_1|$. Similarly, $|^{(n-1)}M_2|=|M_2|$.
Since $M_1^{(n-1)}$ and $^{(n-1)}M_2$ are subsets of $\{0,1\}^{n-1}$,
it follows that $|M_1^{(n-1)}\cup ^{(n-1)}M_2|\leq 2^{n-1}$. Therefore,
$|M_1^{(n-1)}\cap ^{(n-1)}M_2|\geq |M_1|+|M_2|-2^{n-1}$. Let $\mathbf{b}$
denote an arbitrary element of $M_1^{(n-1)}\cap ^{(n-1)}M_2$. It now 
follows that there exist $a,c \in \{0,1\}$ such that $a\mathbf{b} \in M_1$ and
$\mathbf{b}c \in M_2$. Hence, $a\mathbf{b}c \in M_1\otimes M_2$. Thus,
$|M_1\otimes M_2|\geq |M_1|+|M_2|-2^{n-1}$. This completes the proof. 

{\bf Theorem 1:} If $S(\mathbf{v}_n)\leq \frac{5}{8}$, then there exists
a fix-free code $C(\mathbf{v}_n)$.

{\bf Proof:} Clearly, it suffices to prove that $S(\mathbf{v}_n)=\frac{5}{8}$,
implies the existence of a fix-free code $C(\mathbf{v}_n)$. Let us consider three cases.
\begin{enumerate}
\item $k_1=1$
\item $k_1=0,\ k_2=2$
\item $k_1=0,\ k_2\leq 1$
\end{enumerate}

In every case we construct the code $C(\mathbf{v}_n)$ in 
$n$ steps. On step $t$ we add $k_{t}$ words of length $t$ to the code.
The input of step $t$ is a code $C(\mathbf{v}_{t-1})$, the output is a code
$C(\mathbf{v}_{t})$. Thus, on step $n$ we construct $C(\mathbf{v}_n)$.

{\bf Proof of case 1:} We shall now prove that 
$S(\mathbf{v}_n)\leq \frac{3}{4}$ and $k_1=1$ imply the existence of
a fix-free code $C(\mathbf{v}_n)$. This claim is stronger than the 
assertion of the theorem. Put $C(\mathbf{v}_1)=\{0\}$. Suppose that
a fix-free code $C=C(\mathbf{v}_{t-1})$ is constructed; we shall prove 
that on step $t$ we can add $k_t$ words of length $t$ to the code 
without violation of the fix-free property. By lemma~1, it is sufficient
to prove that $|\overrightarrow{F}(C)\otimes \overleftarrow{F}(C)|\geq k_t$.
Put $\delta=S(\mathbf{v}_{t-1})$. Using~(\ref{s_def}),
we get $\delta+\frac{k_t}{2^t}\leq \frac{3}{4}$. Hence,
\begin{equation}\label{case_1_k_t}
k_t\leq 3*2^{t-2}-\delta*2^t.
\end{equation}
Now note that since $0 \in C(\mathbf{v}_{t-1})$, it follows that 
$^0\overrightarrow {F}(C)= \overleftarrow {F}^0(C)=\emptyset$. Therefore,
$\overrightarrow{F}(C)$ is right regular and $\overleftarrow{F}(C)$ is left
regular. It can be easily checked that $|\overrightarrow{F}(C)|=
|\overleftarrow{F}(C)|=2^{t-1}(1-\delta)$. The application of lemma 2
yields 
\begin{equation}\label{case_1_MM}
|\overrightarrow{F}(C)\otimes \overleftarrow{F}(C)|\geq 3*2^{t-2}-\delta*2^t.
\end{equation}
Combining~(\ref{case_1_k_t}) and~(\ref{case_1_MM}), we obtain
$|\overrightarrow{F}(C)\otimes \overleftarrow{F}(C)|\geq k_t$.
This completes the proof of the first case of Theorem 1.

{\bf Proof of case 2:}  We shall now prove that 
$S(\mathbf{v}_n)\leq \frac{3}{4}$,\ $k_1=0$ and $k_2=2$ imply the existence of
a fix-free code $C(\mathbf{v}_n)$. Again, our claim is stronger than the 
assertion of the theorem. Put $C(\mathbf{v}_2)=\{00,11\}$. Suppose that
a fix-free code $C=C(\mathbf{v}_{t-1})$ is constructed; we shall prove that 
$|\overrightarrow{F}(C)\otimes \overleftarrow{F}(C)|\geq k_t$. It is sufficient 
to prove that both inequalities (\ref{case_1_k_t}) and (\ref{case_1_MM}) 
are fulfilled. The proof  of inequality  (\ref{case_1_k_t}) is exactly the same
as above, so we proceed to inequality (\ref{case_1_MM}).

Let us show that $\overrightarrow{F}(C)$ is right regular. 
Assume the converse. Then there exists a vector $\mathbf{b} \in \{0,1\}^{t-2}$
such that both words $0\mathbf{b}$ and $1\mathbf{b}$ are prefix free over
$C(\mathbf{v}_{t-1})$. Let us consider the two cases $^1\mathbf{b}=0$ and
$^1\mathbf{b}=1$ separately. In the first case $0\mathbf{b}$ is prefixed by the 
codeword $00$. In the second case $1\mathbf{b}$ is prefixed by the codeword $11$.
Thus, we have come to a contradiction. 
By the same argument, $\overleftarrow{F}(C)$ is left regular. 
As above, $|\overrightarrow{F}(C)|=|\overleftarrow{F}(C)|=2^{t-1}(1-\delta)$.
The application of lemma 2 yields (\ref{case_1_MM}).
This completes the proof of the second case of Theorem 1.

{\bf Proof of case 3:} 
Since $k_1=0$ and $k_2\leq 1$, it follows that the vector $\mathbf{v}_n$
can be uniquely represented as a sum of four vectors 
$\mathbf{v}_n^1,\mathbf{v}_n^2,\mathbf{v}_n^3,\mathbf{v}_n^4$ 
such that

\begin{equation}\label{v_def}
\left\{
\begin{array}{l}
\mathbf{v}_n^i=\{k_1^i,\ldots,k_n^i\}, \quad $i=1,2,3,4$, \\
S(\mathbf{v}_n^1)=\frac{1}{4},  \\
S(\mathbf{v}_n^2)=S(\mathbf{v}_n^3)=S(\mathbf{v}_n^4)=\frac{1}{8}, \\
\mbox{If } k_t^i\neq 0, \mbox{ then } \forall i^\prime>i, t^\prime< t  
           \quad k_{t^\prime}^{i^\prime}=0. \\
\end{array}
\right.
\end{equation}

Consider the following example of such representation.
$$
\begin{array}{lc}
\mathbf{v}_n   = \{0,0,2,1,2,6,20\} &  S(\mathbf{v}_n)=\frac{5}{8}, \\
\mathbf{v}_n^1 = \{0,0,2,0,0,0,0  \} &  S(\mathbf{v}_n^1)=\frac{1}{4}, \\
\mathbf{v}_n^2 = \{0,0,0,1,2,0,0  \} &  S(\mathbf{v}_n^2)=\frac{1}{8}, \\
\mathbf{v}_n^3 = \{0,0,0,0,0,6,4  \} &  S(\mathbf{v}_n^3)=\frac{1}{8}, \\
\mathbf{v}_n^4 = \{0,0,0,0,0,0,16\} &  S(\mathbf{v}_n^4)=\frac{1}{8}. \\
\end{array}
$$

We shall construct a code $C(\mathbf{v}_n)$ that is a union of four codes 
$ C(\mathbf{v}_n)=C^{00}(\mathbf{v}_n^1)\cup C^{01}(\mathbf{v}_n^2)\cup 
                  C^{10}(\mathbf{v}_n^3)\cup C^{11}(\mathbf{v}_n^4),$
where each code $C^{\alpha\beta}(\mathbf{v}_n^i)$ contains only codewords of the
form $\alpha\star\beta$. 

Thus, for each $t=\overline{1,n}$ the set of codewords of length $t$ is 
composed of $k_t^1$ codewords of the form $0\star 0$,
$k_t^2$ codewords of the form $0\star 1$,
$k_t^3$ codewords of the form $1\star 0$ and
$k_t^4$ codewords of the form $1\star 1$.

We start with an empty code $C(\mathbf{v}_1)=\emptyset$.
Suppose that a fix-free code $C=C(\mathbf{v}_{t-1})$ is constructed;
we shall prove that on step $t$ the code can be extended with 
$k_t^1$ $0\star 0$ codewords,
$k_t^2$ $0\star 1$ codewords,
$k_t^3$ $1\star 0$ codewords and
$k_t^4$ $1\star 1$ codewords of length $t$ 
without violation of the fix-free property. By lemma~1, it is sufficient
to prove that 
\begin{equation}\label{criterion}
\begin{array}{c}
|^0\overrightarrow{F}(C)\otimes \overleftarrow{F}^0(C)|\geq k_t^1, \\
|^0\overrightarrow{F}(C)\otimes \overleftarrow{F}^1(C)|\geq k_t^2, \\
|^1\overrightarrow{F}(C)\otimes \overleftarrow{F}^0(C)|\geq k_t^3, \\
|^1\overrightarrow{F}(C)\otimes \overleftarrow{F}^1(C)|\geq k_t^4. \\
\end{array}
\end{equation}

Put $\delta_i=S(\mathbf{v}_{t-1}^i)$.
Note that, by construction, $\delta_i=0$ and
$\delta_i<S({\mathbf{v}_n^i})$ both imply $\delta_{i+1}=0$.
We shall consider four possible cases:
\begin{enumerate}
\item $\delta_1<\frac{1}{4}, \delta_2=\delta_3=\delta_4=0$
\item $\delta_1=\frac{1}{4}, \delta_2<\frac{1}{8}, \delta_3=\delta_4=0$
\item $\delta_1=\frac{1}{4}, \delta_2=\frac{1}{8}, \delta_3<\frac{1}{8}, \delta_4=0$
\item $\delta_1=\frac{1}{4}, \delta_2=\frac{1}{8}, \delta_3=\frac{1}{8}, \delta_4<\frac{1}{8}$
\end{enumerate}
In all the cases we use the fact that 
\begin{equation}\label{k_t_upper_bound}
k^i_t\leq 2^t(S(\mathbf{v}_n^i)-\delta_i).
\end{equation}

{\bf Case 3.1:} $\delta_1<\frac{1}{4}, \delta_2=\delta_3=\delta_4=0.$
Using~(\ref{k_t_upper_bound}), we get
$$
\begin{array}{cc}
k_t^1\leq 2^{t-2}-\delta_1*2^t, & k_t^2\leq 2^{t-3}, \\
k_t^3\leq 2^{t-3}, & k_t^4\leq 2^{t-3}. \\
\end{array}
$$
It can be easily checked that 
$$
\begin{array}{c}
|^0\overrightarrow {F}(C)|=|\overleftarrow {F}^0(C)|=2^{t-2}-\delta_1*2^{t-1}, \\
|^1\overrightarrow {F}(C)|=|\overleftarrow {F}^1(C)|=2^{t-2}.
\end{array}
$$
The application of lemma~2 yields
$$
\begin{array}{c}
|^0\overrightarrow{F}(C)\otimes \overleftarrow{F}^0(C)|
                                \geq 2^{t-2}-\delta_1*2^t \geq k_t^1, \\
|^0\overrightarrow{F}(C)\otimes \overleftarrow{F}^1(C)|
                 \geq 2^{t-2}-\delta_1*2^{t-1}> 2^{t-3}\geq k_t^2, \\
|^1\overrightarrow{F}(C)\otimes \overleftarrow{F}^0(C)|
                 \geq 2^{t-2}-\delta_1*2^{t-1}> 2^{t-3}\geq k_t^3, \\
|^1\overrightarrow{F}(C)\otimes \overleftarrow{F}^1(C)|\geq 2^{t-2} > k_t^4. \\
\end{array}
$$
This completes the proof of case 3.1.

{\bf Case 3.2:} $\delta_1=\frac{1}{4}, \delta_2<\frac{1}{8}, \delta_3=\delta_4=0.$
By the same argument as above
$$
\begin{array}{cc}
k_t^1=0, & k_t^2\leq 2^{t-3}-\delta_2*2^t, \\
k_t^3\leq 2^{t-3}, & k_t^4\leq 2^{t-3}. \\
\end{array}
$$
We see that 
$$
\begin{array}{c}
|^0\overrightarrow {F}(C)|= 2^{t-2}-(\frac{1}{4}+\delta_2)*2^{t-1},\\
|\overleftarrow {F}^0(C)|=2^{t-3}, \\
|^1\overrightarrow {F}(C)|=2^{t-2} \\
|\overleftarrow {F}^1(C)|=2^{t-2}-\delta_2*2^{t-1}.\\
\end{array}
$$
By lemma~2, we have
$$
\begin{array}{c}
|^0\overrightarrow{F}(C)\otimes \overleftarrow{F}^1(C)|
                 \geq 2^{t-3}-\delta_2*2^t\geq k_t^2, \\
|^1\overrightarrow{F}(C)\otimes \overleftarrow{F}^0(C)|
                                               \geq 2^{t-3} \geq k_t^3, \\
|^1\overrightarrow{F}(C)\otimes \overleftarrow{F}^1(C)|
                 \geq 2^{t-2}-\delta_2*2^{t-1} > 2^{t-3}\geq k_t^4. \\
\end{array}
$$
This completes the proof of case 3.2.

{\bf Case 3.3:} $\delta_1=\frac{1}{4}, \delta_2=\frac{1}{8}, 
                        \delta_3<\frac{1}{8}, \delta_4=0.$
As above, 
$$
\begin{array}{cc}
k_t^1=0, & k_t^2=0, \\
k_t^3\leq 2^{t-3}-\delta_3*2^t, & k_t^4\leq 2^{t-3}. \\
\end{array}
$$
It is easily shown that 
$$
\begin{array}{c}
|\overleftarrow {F}^0(C)|=2^{t-3}-\delta_3*2^{t-1}, \\
|^1\overrightarrow {F}(C)|=2^{t-2}-\delta_3*2^{t-1} \\
|\overleftarrow {F}^1(C)|=3*2^{t-4}.\\
\end{array}
$$
Applying lemma~2, we obtain
$$
\begin{array}{c}
|^1\overrightarrow{F}(C)\otimes \overleftarrow{F}^0(C)|
                           \geq 2^{t-3}-\delta_3*2^t \geq k_t^3, \\
|^1\overrightarrow{F}(C)\otimes \overleftarrow{F}^1(C)|
                 \geq 3*2^{t-4}-\delta_3*2^{t-1} > 2^{t-3}\geq k_t^4. \\
\end{array}
$$
This completes the proof of case 3.3.

{\bf Case 3.4:} $\delta_1=\frac{1}{4}, \delta_2=\frac{1}{8}, 
                        \delta_3=\frac{1}{8}, \delta_4<\frac{1}{8}.$
As above,
$$
\begin{array}{cc}
k_t^1=0, & k_t^2=0, \\
k_t^3=0, & k_t^4\leq 2^{t-3}-\delta_4*2^t. \\
\end{array}
$$
One can easily see that 
$$
\begin{array}{c}
|^1\overrightarrow {F}(C)|=
|\overleftarrow {F}^1(C)|=2^{t-2}-(\frac{1}{8}+\delta_4)*2^{t-1}.\\
\end{array}
$$
By lemma~2, we have
$$
\begin{array}{c}
|^1\overrightarrow{F}(C)\otimes \overleftarrow{F}^1(C)|
                 \geq 2^{t-3}-\delta_4*2^t \geq k_t^4. \\
\end{array}
$$
This completes the proof of the theorem.

\section{Upper Bound for the Redundancy}

{\bf Theorem 2:} For each probability distribution $p=\{p_1,\ldots,p_m\}$
there exists a binary fix-free code $C$ where the average length of the 
codewords $L(C)$ satisfies
$$
L(C)< H(p)+4-\log_{2}{5}.
$$
{\bf Proof:} By $l_1,\ldots,l_m$ denote the codeword lengths. We define 
$$l_i=\left\lceil -\log_{2}{p_i}+3-\log_{2}5\right\rceil.$$ 
It follows that
$$
\sum\limits_{i=1}^{m}2^{-l_i}\leq 
\sum\limits_{i=1}^{m}2^{\log_{2}p_i-3+\log_{2}5}=
\frac{5}{8}\sum\limits_{i=1}^{m}p_i=\frac{5}{8}.
$$
By theorem~1 there exists a fix-free code $C$ with the codeword lengths
$l_1,\ldots,l_m$. The average length of this code~is
$$
\begin{array}{c}
L(C)=\sum\limits_{i=1}^{m}p_i*l_i< 
     \sum\limits_{i=1}^{m}p_i(-\log_{2}p_i+4-\log_{2}5)= \\
     H(p)+(4-\log_{2}5)\sum\limits_{i=1}^{m}p_i=H(p)+4-\log_{2}5. \\
\end{array}
$$
This completes the proof.

\end{document}